\documentclass[aps,twocolumn,showpacs,preprintnumbers,noshowpacs,superscriptaddress,6pt]{revtex4-2}

\usepackage{graphicx}  % Include figure files
\usepackage{subfigure}
\usepackage{multirow}
\usepackage{adjustbox}
\usepackage{natbib}

\usepackage{fancyhdr}
\usepackage{longtable}
\usepackage{parskip}
\usepackage[T1]{fontenc}
\usepackage{dcolumn}   % Align table columns on decimal point

\usepackage{bm}        % bold math
\usepackage{amsfonts}  % Common math fonts
\usepackage{amsmath}   % Common math functions
\usepackage{amssymb}   % Common math symbols
\usepackage{siunitx}
\usepackage{booktabs}
\usepackage{physics}
\usepackage{mathtools}
\usepackage{flafter}
\usepackage{threeparttable}
\usepackage{subfiles} % Best loaded last in the preamble
\usepackage{subfigure}
\usepackage{import}
\usepackage{float}

\usepackage{diagbox}

\usepackage{ulem}

\usepackage{colortbl}
\usepackage{color}
\usepackage{ulem}
\definecolor{BLUE}{rgb}{0.0,0.0,1.0}

\begin{document}

\title{Multiple-ionization energy difference of~$^{163}$Ho and~$^{163}$Dy atoms}

\author{I.~M.~Savelyev}
\affiliation{Department of Physics, St. Petersburg State University, 7/9 Universitetskaya nab., 199034 St. Petersburg, Russia}

\author{M.~Y.~Kaygorodov}
\affiliation{Department of Physics, St. Petersburg State University, 7/9 Universitetskaya nab., 199034 St. Petersburg, Russia}

\author{Y.~S.~Kozhedub}
\affiliation{Department of Physics, St. Petersburg State University, 7/9 Universitetskaya nab., 199034 St. Petersburg, Russia}

\author{I.~I.~Tupitsyn}
\affiliation{Department of Physics, St. Petersburg State University, 7/9 Universitetskaya nab., 199034 St. Petersburg, Russia}

\author{V.~M.~Shabaev}
\affiliation{Department of Physics, St. Petersburg State University, 7/9 Universitetskaya nab., 199034 St. Petersburg, Russia}

\date{\today}
\begin{abstract}
The multiple-ionization energy difference of~$^{163}$Ho and~$^{163}$Dy atoms is evaluated for the ionization degree~$q=30,48,56$. 
The calculations are performed by means of the large-scale relativistic configuration interaction method  combined with the many-body perturbation theory.
The quantum electrodynamics, nuclear recoil, and frequency-dependent Breit interaction corrections are taken into account.
The obtained theoretical values are within 1 eV uncertainty.
These results can help to increase accuracy of the laboratory neutrino mass limit, provided they are accompanied by the corresponding experiment on electron capture in Ho and a precise measurement of the ion mass difference.
\end{abstract}

\maketitle

\section{Introduction}
One of the most challenging problem of the modern physics is the determination of the neutrino mass absolute value~\cite{2003_KingS_RepProgPhys, 2013_DrexlinG_AdvHighEnergyPhys, 2021_FormaggioJ_PhysicsReports914}.
Presently, the upper limit for the sum of the neutrino masses obtained from the analysis of astrophysical data is tenth of eV~\cite{2018_PDG_PhysRevD, 2020_AghanimN_AA}.
The laboratory limits on the mass of the electronic antineutrino, obtained through kinematics studies of the beta-decay process in tritium, are at about 1 eV level~\cite{2005_KrausC_EurPhysJC, 2011_AseevV_PhysRevD, 2019_KATRIN_PhysRevLett123}, or even less than $0.8$~eV, obtained recently at the KATRIN installation~\cite{2021_AkerM_ArXiv210508533}. 
According to the Standard Model which incorporates the CPT symmetry, the neutrino mass must be exactly equal to the antineutrino one.
However, the direct experiments on establishing the limits on the neutrino masses are also of undoubted interest, especially if they can achieve a level of accuracy comparable to the laboratory antineutrino limit.
\par
Nowadays, the best direct laboratory limit on the mass of the electron neutrino is about 225 eV~\cite{1987_SpringerP_PhysRevA}.
To substantially decrease the current limit, an experiment to study electron capture (EC) process in~$^{163}\textrm{Ho}$ was proposed~\cite{2007_KlugeH_NuclPhysNews, 2015_AlpertB_EurPhysJC, 2017_GastaldoL_EurPhysJSpecTop}.
%\par
In the EC process, a bound electron is captured by a proton in the Ho nucleus.
The proton transforms into a neutron and an electronic neutrino, the atom is left in an excited state, and the energy~$Q$, referred to as the mass excess, is released: 
\begin{align}
    ^{163}\mathrm{Ho}\xrightarrow{\mathrm{EC}} \mathrm{^{163}Dy^*}+\nu_e.
\end{align}
The energy~$Q$ is shared between the emitted neutrino and the atomic excitation of the daughter Dy atom.
The experiment concept is to perform a calorimetric measurement of all the atomic de-excitation spectrum except the energy removed by the electronic neutrino.
The nonzero neutrino mass affects the shape of the spectrum.
Fitting the spectrum and analyzing its high-energy end-point parts, one can obtain information about the neutrino mass.
However, to establish the limit on the neutrino mass in the EC process, it is necessary to know the mass excess~$Q$, \textit{i.e.}, the mass difference between~$^{163}\textrm{Ho}$ and~$^{163}\textrm{Dy}$ atoms, at the appropriate level of accuracy.
\par
At present, the most accurate mass determination is available in the experiments with ions in Penning-trap mass spectrometers~\cite{2015_EliseevS_PhysRevLett, 2020_RischkaA_PhysRevLett, 2021_FilianinP_PhysRevLett127}.
To recalculate the mass difference of the ions into the mass difference of the neutral atoms, one has to perform theoretical evaluations of the binding energy of the ionized electrons.
The aim of the present work is to consider various ionization degrees of the Ho and Dy species, masses of which are to be measured, and to evaluate the related binding energy differences to restore the atomic mass difference from the ionic one.
\par
To perform the calculations required, we use the configuration interaction (CI) method in the basis of Dirac-Fock-Sturm orbitals combined with the many-body perturbation theory which is implemented in CI-DFS program~\cite{2003_TupitsynI_PhysRevA, 2005_TupitsynI_PhysRevA} and Kramers-restricted configuration interaction module KR-CI~\cite{2010_KnechtS_JChemPhys} of the DIRAC19~\cite{DIRAC19} program.
It is essential to note that the current implementations of the multireference coupled cluster approach, which is very effective for highly correlated systems, can not be applied to the Ho and Dy atoms due to their complex electron configuration.
The quantum-electrodynamics (QED) corrections to the considered binding-energy differences are evaluated within the model QED operator approach.
\par
The paper is organized as follows: in Sec. \ref{sec:2} the multiple-ionization energy difference is introduced and some features of the CI-DFS and the KR-CI methods are given; the numerical details are presented in Sec. \ref{sec:3} together with the discussion of the results; Sec. \ref{sec:4} concludes the paper.
\par
The atomic units are used throughout the paper.

\section{Theory}\label{sec:2}
\subsection{Ho and Dy atomic mass}\label{subsec:2:a}
We consider the mass difference~$\Delta m^q$ of the~$^{163}$Ho$^{q+}$ and~$^{163}$Dy$^{q+}$ ions with the same ionization degree~$q$,
\begin{equation}
    \Delta m^q = \Delta m_n + m_e + \Delta E^q,
\end{equation}
where~$\Delta m_n$ is the mass difference between the~$^{163}$Ho and~$^{163}$Dy nuclei,~$\Delta E^{q}$ is the difference between the total electronic binding energies of the Ho and Dy ions.
The case~$q=0$ corresponds to the neutral Ho and Dy atoms mass difference
\begin{equation}
    \Delta m^0 = \Delta m_n + m_e + \Delta E^0.
\end{equation}
The ionic mass difference~$\Delta m^q$ is related to the atomic one~$\Delta m^0$ through the atomic and ionic binding energy differences
\begin{equation}
    \Delta m^0 = \Delta m^q + \Delta E^{0,q},
\end{equation}
where the notation for the \textit{secondary} difference of the binding energies,
\begin{equation}
     \Delta E^{0,q} = \Delta E^0- \Delta E^q,
\end{equation}
is introduced.
The quantity~$\Delta E^{0,q}$ is interpreted as the difference between the binding energies of the outermost~$q$ electrons in the Ho and Dy atoms.

\subsection{Method of calculation}\label{subsec:2:b}
\subsubsection{CI-DFS method}
In the present work, the primary method for evaluation of the binding-energy differences~$\Delta E^{0,q}$ is the relativistic large-scale CI method combined with the many-body perturbation theory in the basis of the Dirac-Fock-Sturm orbitals (CI-DFS)~\cite{2003_TupitsynI_PhysRevA, 2005_TupitsynI_PhysRevA}.
The method is based on the finding of the eigenvalues of the Dirac-Coulomb-Breit (DCB) Hamiltonian represented in the many-electron basis.
The DCB Hamiltonian for an~$N$-electron atom has the form
\begin{equation}\label{eq:H_DCB}
H_{\mathrm{DCB}} = \Lambda^+ (H_{\mathrm{D}} + H_{\mathrm{C}}+ H_{\mathrm{B}}) \Lambda^+,
\end{equation}
where~$H_{\mathrm{D}}$ is the sum of the one-electron Dirac Hamiltonians,
\begin{equation}
    H_{\mathrm{D}} = \sum_{i=1}^N \left[ (\bm{\alpha}_i \cdot \bm{p}_i)c+(\beta-1)mc^2+V(r_i) \right],
\end{equation}
$H_{\mathrm{C}}$ and~$H_{\mathrm{B}}$ are the Coulomb and Breit electron-electron interaction operators, respectively,
\begin{equation}
    H_{\mathrm{C}} = \frac{1}{2}\sum_{i\neq j}^N \frac{1}{r_{ij}},
\end{equation}
\begin{equation}
    H_{\mathrm{B}} = -\frac{1}{2} \sum_{i \neq j}^N \frac{1}{2r_{ij}}\Big[\bm{\alpha}_i\cdot\bm{\alpha}_j+\frac{(\bm{\alpha}_i\cdot\bm{r}_{ij})(\bm{\alpha}_j\cdot\bm{r}_{ij})}{r_{ij}^2}\Big].
\end{equation}
Here~$\bm{\alpha}$ is a vector incorporating the Dirac matrices,~$\bm{p}$ is the momentum operator,~$\bm{r}_{ij}$ is the position of the~$i$-th electron relative to the~$j$-th one, and~$r_{ij}=|\bm{r}_{ij}|$.
The Fermi model for the nuclear-charge density with the standard parameters is used to represent the nuclear potential~$V(r)$ in the Hamiltonian~$H_{\mathrm{D}}$.
The operators~$\Lambda^+$ in Eq. (\ref{eq:H_DCB}) project the Hamiltonian on the positive-energy spectrum of the Dirac-Fock operator~$h_{\mathrm{DF}}$.
\par
The many-electron basis in the CI-DFS method consists of the symmetry adapted linear combinations of the Slater determinants referred to as configuration state functions (CSFs).
The active space is spanned by the eigenfunctions of the~$h_{\mathrm{DF}}$ Hamiltonian operator in the basis of the occupied Dirac-Fock~$\varphi^{\mathrm{DF}}$ and the virtual Dirac-Fock-Sturm~$\varphi^{\mathrm{DFS}}$ one-particle orbitals, which are obtained as numerical solutions of the DF and DFS equations, respectively.
For some~$\varphi_i^{\mathrm{DFS}}$ orbital, the DFS equation contains an adjustable parameter~$\varepsilon_i$, which is chosen to be one of the occupied DF orbital energy and related to the spatial form of the DFS orbital.
The corresponding DF orbital is termed as reference one.
Finally, the configuration space is generated in accordance with the restricted active space (RAS) scheme~\cite{1988_OlsenJ_JChemPhys}.
\par
When the number of configurations becomes too large for the direct employment of the CI-DFS method, one can take into account a contribution from some other configurations using the perturbation theory (PT).
Within the combined approach the whole configuration space is divided into two subspaces: the model space~$P$ and its orthogonal complementary~$Q$.
In the~$P$ space the correlations are treated utilizing the CI method whereas the correlations with the~$Q$ space are treated perturbatively.
This approach allows one to take into account the contribution with the~$Q$-space configurations at cheaper price than in the pure CI method, although only approximately. 
\par
The optimal scheme of choosing the~$P$ and~$Q$ spaces, \textit{a~priori}, is not known.
In the present work, an automatic procedure based on identification and sorting of the most important configurations using the perturbation theory of the second order is applied.
The user provided number of configurations~$N_p$, which have the highest weight, form the model space~$P$, the rest of the configurations belong to the~$Q$ space.
Varying the parameter~$N_p$, for a given configuration space one can find the desired balance between the required precision and the computational cost.
\par
The QED correction to the binding-energy difference~$\Delta E^{0,q}$ is calculated using the model QED operator approach.
Within this approach the model QED operator~$V^{\mathrm{QED}}_{\mathrm{mod}}$~\cite{2013_ShabaevV_PhysRevA, 2015_ShabaevV_CompPhysComm, *2018_ShabaevV_CompPhysComm} is incorporated into the many-electron Hamiltonian~$H_{\mathrm{DCB}}$~(\ref{eq:H_DCB}) as well as into the one-electron DF Hamiltonian~$h
_{\mathrm{DF}}$.
The frequency-dependent Breit interaction correction is calculated as the expectation value of the full frequency-dependent electron-electron interaction operator in the Coulomb gauge with the correlated many-electron wave function.
The nuclear recoil correction is evaluated as the expectation value of the relativistic mass shift operator~\cite{Shabaev1985TMP_Recoil, *Shabaev1988, Palmer1987, 1998_ShabaevV_PhysRevA}.

\subsubsection{KR-CI module}
In order to additionally check the relevance of tacking into account the correlation effects by the CI-DFS approach, we use a different implementation of the CI method.
To this end, the KR-CI module~\cite{2010_KnechtS_JChemPhys}, which is a part of the DIRAC program~\cite{DIRAC19} is used.
The module employs the Dirac-Coulomb Hamiltonian
\begin{equation}\label{eq:H_DC}
    H_{\mathrm{DC}} = \Lambda^+ (H_{\mathrm{D}} + H_{\mathrm{C}})\Lambda^+,
\end{equation}
where the~$\Lambda^+$ operators, as it is in Eq.~(\ref{eq:H_DCB}), project out the negative energy solutions of the DF Hamiltonian.
The occupied and virtual one-electron basis functions are obtained as solutions of the Dirac-Fock equations in the finite basis set of the primitive Gaussian functions.
The many-electron functions in the KR-CI method are the Slater determinants generated from the active DF orbitals.
It should be noted, that the DIRAC package, and the KR-CI module in particular, is a molecular-oriented program and does not take advantage of the atomic center-field symmetry.
Therefore, it is much more difficult to apply this program for atomic calculations, compared to the CI-DFS method with the configuration space of the same size.

\section{Numerical results and discussion}\label{sec:3}
To estimate the energy~$\Delta E^{0,q}$, first, the ground-state energy difference of the neutral Ho and Dy atoms,~$\Delta E^0$, is considered. 
The electron configuration of the ground state of holmium is ~$[\mathrm{Xe}] 4f^{11} 6s^2$, term with~$J=15/2$.
Dysprosium has one less electron on the~$f$ shell.
Its configuration of the ground state is~$[\mathrm{Xe}] 4f^{10} 6\mathrm{s}^2$, term with~$J=8$.
\par
To evaluate the binding-energy difference for the ions,~$\Delta E^q$, we chose the ionization degrees~$q=30,48,56$, since the corresponding configurations simplify the calculations.
These ionization degrees correspond to the following configurations of the Ho-Dy pair of ions~\cite{2004_RodriguesG_ADNDT}:~$[\mathrm{Kr}]4d^1$ and~$[\mathrm{Kr}]$ for~$q=30$,~$[\mathrm{Ar}]3d^1$ and~$[\mathrm{Ar}]$ for~$q=48$,~$[\mathrm{Ne}]3s^1$ and~$[\mathrm{Ne}]$ for~$q=56$, respectively.
\par
The first calculation data set of Ho and Dy binding-energy difference is presented in Table~{\ref{table:1}}.
The results are obtained with the~$H_{\mathrm{DC}}$ Hamiltonian, the single (S) and double (D) excitations are taken into account.
The electrons from valence~$4f$ and~$6s$ orbitals are considered as active ones, while basis set dependence on the number of virtual orbitals is studied. 
One reads the table as follows: the cell with, e.g.,~$n=2$,~$l=3$ means that~$2$ additional virtual orbitals for each symmetry up to the~$l=3$ shell ($7s8s6p7p5d6d5f6f$) are added into the  active space.
The notation is applied to the Ho and Dy atoms simultaneously in order to provide similar description of the one-electron basis sets for both atoms.
Extrapolating the results to the complete one-electron basis set limit~$(n\to\infty,l\to\infty)$, one obtains the value for the neutral Ho and Dy ground state energy difference,~$\Delta E^0_{\mathrm{DC}}=-459.892(5)$ a.u., where the uncertainty has a pure numerical origin.
The calculations are carried out using the CI-DFS method combined with PT.
Meanwhile, we keep under control the error associated with the usage of the PT in the~$Q$ space.
Using smaller numbers of active orbitals, we compare the CI-DFS combined with PT results with the pure CI ones and ensure that the difference between them is less than~$10^{-4}$ a.u.

\begin{table}[H]
\centering

\caption{The dependence of the ground-states energy difference for the neutral Ho and Dy atoms,~$\Delta E^0_{\mathrm{DC}}$, computed with the CI-DFS method on the number of virtual orbitals~$n$ with orbital quantum number up to~$l$ included into the active space~(a.u.).
The energy difference is evaluated with the Coulomb interaction operator only and the SD excitations from the~$4f6s$ shells are considered.}
\label{table:1}
\begin{tabular}{S[table-format=1.1] S[table-format=4.6] S[table-format=4.6] S[table-format=4.6] S[table-format=4.6]}

\toprule
\multicolumn{1}{c}{\backslashbox{$l$}{$n$}}  &
\multicolumn{1}{c}{$1$}  &
\multicolumn{1}{c}{$2$}  &
\multicolumn{1}{c}{$3$}  &
\multicolumn{1}{c}{$4$}  \\
\midrule
0 & -459.80897 & -459.80897 & -459.80897 & -459.80898 \\
1 & -459.80907 & -459.80924 & -459.80931 & -459.80927 \\
2 & -459.81402 & -459.81519 & -459.81549 & -459.81548 \\
3 & -459.84025 & -459.84887 & -459.85170 & -459.87742 \\
4 & -459.85249 & -459.87043 & -459.85245 & -459.87980 \\
5 & -459.85364 & -459.87340 & -459.88166 & -459.88456 \\
6 & -459.85465 & -459.87525 & -459.88366 & -459.88703 \\
7 & -459.85497 & -459.87592 & -459.88448 & -459.88797 \\
\bottomrule

\end{tabular}
\end{table}

To additionally verify our results, we calculate the same quantity~$\Delta E^0_{\mathrm{DC}}$ with a different implementation of the CI method.
For this purpose we employ the KR-CI module, which is a part of the DIRAC package.
The Dyall's \texttt{ae4z} basis set~\cite{2010_GomesA_TheorChemAcc} was employed in the calculations.
The  orbitals with active electrons in the KR-CI calculations were the same as in the CI-DFS ones, but the virtual orbitals were different.
Analyzing the convergence of the KR-CI results with respect to the number of the virtual orbitals in a similar way as we did in the CI-DFS calculations, we obtained the value~$\Delta E^0_{\mathrm{DC}}=-459.887(8)$ a.u.
Both the CI-DFS and KR-CI results are in agreement with each other within the estimated uncertainty.
\par
Then we proceed with the calculations of the part of the energy difference~$\Delta E^0$, which accounts for the core-core and core-valence correlations.
To this end, we artificially separate electrons into the valence, active-core, and frozen-core ones and calculate the active-core interaction correction by means of the PT.
Initially, the active electrons were~$4f6s$, while the remaining were assigned to the frozen core.
Then, we gradually unfreeze the electrons, starting from the orbital~$5p$, add them to the active-core ones, and compute the correction from the correlations of the active core orbitals to the Ho-Dy ground state energy difference.
Following this procedure, we found that the contribution due to inclusion of the~$4s4p$ electrons into the correlation problem is several times smaller than that caused by inclusion of the~$5s5p$ electrons, and, therefore, we stopped the further expansion of the occupied active space.
We supplement the corrections associated with the core correlations with a larger uncertainty than that we obtained from the convergence analysis.
Overall, the~$4s4p4d5s5p4f6s$ electrons have been correlated; the contribution from the correlation of the~$4s4p4d5s5p$ electrons provides the main source of the numerical uncertainty for the neutral Ho and Dy ground-state energy difference.
\par
We also calculated various corrections to the~$\Delta E^0_{\mathrm{DC}}$ value.
First, we computed the correction from the inclusion of the Breit interaction operator~$H_{\mathrm{B}}$ into the many-electron DC Hamiltonian~(\ref{eq:H_DC}).
The Davidson correction was evaluated using a simple method proposed in Ref.~\cite{1974_LanghoffS_IntJQuantumChem}.
The QED corrections were evaluated within the model QED operator approach according to the procedure described in Sec. \ref{subsec:2:b}.
The Breit frequency-dependent interaction correction was calculated as the expectation value of the frequency-dependent part of the electron-electron interaction operator in the Coulomb gauge with the correlated many-electron wavefunction.
Lastly, the nuclear recoil correction within the Breit approximation was evaluated.
\par
In the calculations of the ground-state energy difference~$\Delta E^q$ for Ho and Dy ions, we performed the same methodological analysis of each contribution as for the neutral atom case.
Specifically, first we computed the~$\Delta E^q_{\mathrm{DC}}$ value treating the following electrons as active ones:~$2s2p3s$ for the ionization degree~$q=56$,~$3s3p3d$ for~$q=48$, and~$4s4p4d$ for~$q=30$; the remaining electrons were considered as the frozen core.
Further, the corrections from the Breit interaction, the Davidson correction, the QED, frequency-dependent Breit interaction, and nuclear recoil corrections were evaluated for each pairs of ions.
Lastly, all the electrons, which have been frozen on the previous stage, are considered now as active ones and the corrections from the interaction with these active core electrons are computed for each pair of ions using PT.
\par
The results for the ground-state energy difference~$\Delta E^0$ of Ho and Dy atoms as well as the results for the ground-state energy difference~$\Delta E^q$ of the related ions with the ionization degree~$q=30,48,56$ are collected in Table~\ref{table:2}.
For each~$q$ we combine the contributions obtained within the framework of the DCB Hamiltonian into the single quantity and place it into the second column of the table. 
The contributions from the frequency-dependent Breit interaction and the QED effects are presented in the next two columns.
The last column represents the sum of all the contributions.
For the total values, the first parentheses represent the numerical uncertainty, the uncertainty associated with the finite nuclear size is given in the second parentheses.
\onecolumngrid\
\begin{table}[H]
\centering

\caption{The contributions to the ground-state energy difference~$\Delta E^q$ computed with the CI-DFS method for the neutral Ho and Dy atoms and their ions from the effects within the Breit approximation,~$\Delta E^q_{\mathrm{DCB}}$, from the Breit frequency-dependent interaction correction,~$\Delta E^q_{\mathrm{BRFD}}$, and the QED effects,~$\Delta E^q_{\mathrm{QED}}$.
For the total value~$\Delta E^q$, the numbers in the first parentheses represent the numerical accuracy, whereas numbers in the second parentheses represent the uncertainty associated with the finite nuclear size.}
\label{table:2}
\begin{tabular}{l S[table-format=-5.6] S[table-format=-4.8] S[table-format=4.8] S[table-format=-3.5(5),table-align-text-post=false] S[table-format=-5.5(5),table-align-text-post=false]}

\toprule
\multicolumn{1}{c}{$q$}  & \multicolumn{1}{c}{$\Delta E^q_{\mathrm{DCB}}$ (a.u.) } & \multicolumn{1}{c}{$\Delta E^q_{\mathrm{BRFD}}$ (a.u.)} & \multicolumn{1}{c}{$\Delta E^q_{\mathrm{QED}}$ (a.u.)} & \multicolumn{1}{c}{$\Delta E^q$ (a.u.)} & \multicolumn{1}{c}{$\Delta E^q$ (eV)}\\
\midrule
0  & -459.294 & -0.015 & 0.411 & -458.898(26)$(13)$ & -12487.25(71)$(35)$ \\
30 & -458.803 & -0.015 & 0.410 & -458.408(19)$(13)$ & -12473.91(52)$(35)$ \\
48 & -493.115 & -0.014 & 0.402 & -492.727(9)$(13)$  & -13407.79(12)$(35)$ \\
56 & -494.101 & -0.010 & 0.515 & -493.596(6)$(13)$  & -13431.44(17)$(35)$ \\

\bottomrule

\end{tabular}
\end{table}
\twocolumngrid\

Consider now the desired secondary difference~$\Delta E^{0,q}$, which corresponds to the multiple-ionization energy difference of the outermost~$q$ electrons in the Ho and Dy atoms.
For the ionization degree~$q=30,48,56$, this quantity is presented in Table~\ref{table:3}.
Again, we present the contributions from the QED effects and the frequency-dependent Breit interaction in columns~$\Delta E^q_{\mathrm{QED}}$, and~$\Delta E^q_{\mathrm{BRFD}}$, respectively.
All the other contributions are incorporated into the quantity~$\Delta E^q_{\mathrm{DCB}}$.
\begin{table}[H]
\centering

\caption{The contributions to the multiple ionization energy difference of the outermost~$q$ electrons in the neutral Ho and Dy atoms,~$\Delta E^{0,q}$, from the effects within the Breit approximation,~$\Delta E^{0,q}_{\mathrm{DCB}}$, the Breit frequency-dependent interaction correction~$\Delta E^{0,q}_{\mathrm{BRFD}}$, and the QED effects,~$\Delta E^{0,q}_{\mathrm{QED}}$ (eV).}
\label{table:3}
\begin{tabular}{l S[table-format=4.1] S[table-format=-1.2] S[table-format=-1.2] S[table-format=3.1(1)]}

\toprule
\multicolumn{1}{c}{$q$}  &
\multicolumn{1}{c}{$\Delta E^{0,q}_{\mathrm{DCB}}$}  &
\multicolumn{1}{c}{$\Delta E^{0,q}_{\mathrm{BRFD}}$}  &
\multicolumn{1}{c}{$\Delta E^{0,q}_{\mathrm{QED}}$}  &
\multicolumn{1}{c}{$\Delta E^{0,q}$}  \\
\midrule
30  & -13.3 &  0.00 & 0.03 & -13.3(9) \\
48  & 920.3 & -0.03 & 0.26 & 920.5(7) \\
56  & 947.2 & -0.14 & -2.83 & 944.2(7) \\
\bottomrule

\end{tabular}
\end{table}
\par
In the secondary difference~$\Delta E^{0,q}$ the uncertainty associated with the finite nuclear size cancels out and the~$\Delta E^{0,q}$ quantity turns out to be more stable with respect to variation of the nuclear parameters than the individual terms.
Also, for~$\Delta E^{0,q}$ we obtain large cancellation of the QED and Breit frequency-dependent interaction corrections for~$q=30$ and~$q=48$ degrees of ionization.
At the same time, for~$q=56$ the QED contribution is about~$-3$ eV.
This is due to the fact that the valence~$3s$ electron is present in~$\mathrm{Ho}^{56+}$ configuration, whereas for~$\mathrm{Ho}^{48+}$ and for~$\mathrm{Ho}^{30+}$ ions the~$3d$ and~$4d$ electrons, respectively, are valence.
The frequency-dependent Breit interaction correction to the energy difference varies from about~$1$ to~$5$\% of the~$\Delta E^{0,q}_{\mathrm{QED}}$ value, being larger for higher degrees of ionization~$q$.
For all ionization degrees~$q$ considered, the recoil correction to~$\Delta E^{0,q}$ turns out to be two orders of magnitude smaller than the QED correction.
The corrections that are not canceled in the secondary ground-state energy difference~$\Delta E^{0,q}$ are mainly due to the inter-electronic interaction effects within the Breit approximation.
The Davidson correction to~$\Delta E^{0,q}$ amounts to about~$-0.1$ eV for all considered values of~$q$.
The main uncertainty in the secondary difference~$\Delta E^{0,q}$ comes from the correlation of the electrons which are present in the neutral atoms but absent in the considered ions, namely,~$4s4p4d5s5p$ electrons.

\section{Conclusion}\label{sec:4}
In the planing experiments on lowering the electronic neutrino upper mass limit through the analysis of the electron capture process in~$^{163}$Ho it is necessary to know the mass difference of the neutral Ho and Dy atoms with high accuracy. 
This value can be determined via the mass difference of the related ions, which can be measured experimentally with high precision, and the multiple-ionization energy difference. 
The last value has been precisely evaluated in the present work for the ionization of~$q=30,48,56$ outermost electrons. 
The calculations are carried out using the large-scale relativistic configuration interaction method combined with many-body perturbation theory.
The QED, nuclear recoil, and frequency-dependent Breit interaction corrections are taken into account.
The obtained results for the multiple-ionization energy difference are within~$1$~eV theoretical uncertainty.
The corresponding calculations for other ionization degrees, can be performed within the developed approach if required.
\section{Acknowledgments}
We thank H.-J. Kluge and Y. N. Novikov for stimulating discussions.
This work was supported by RFBR and ROSATOM according to the research project No.~20-21-00098.

\bibliographystyle{apsrev}
\bibliography{main}
 \end{document}